\documentclass{rnaastex}

\begin{document}

\title{Confirmation of Intervening Filaments of Galaxies at the Redshifts of {\it Chandra} WHIM Absorption Features}

\correspondingauthor{Shea Brown}
\email{shea-brown@uiowa.edu}

\author{Shea Brown}
\author{Philip Kaaret}
\author{Anna Zajczyk}
\affiliation{Department of Physics \& Astronomy \\
The University of Iowa \\
Iowa City, Iowa 52245, USA}


\section{} 
We report the {\it a posteriori} confirmation of two large-scale filaments along the sight-line of the blazar 1ES~1553+113, which correspond to Warm-Hot Intergalactic Medium \citep[WHIM;][]{cen99,dave01} absorption features in the X-ray and far ultraviolet. \cite{nica13} reported 5 low signal-to-noise absorption features along the sight line to the $z \geq 0.4$ blazar 1ES 1553+113. These CV and CVI K$\alpha$ absorption lines correspond to potential WHIM systems at $z_X = 0.312, z_X = 0.237$, and a cluster of lines with an average redshift of $< z_X > = 0.133$. The two highest redshift systems also have corresponding BLA lines detected in the {\it FUV}. Unlike the detections of \cite{buot09} and \cite{fang10}, where there was an obvious low redshift intervening structure (the Sculpor Wall at $z\approx 0.03$)\footnote{The claims of WHIM absorption at the Sculptor Wall has been challenged by \cite{nica16}.}, the redshifts of \cite{nica13} are at distances where spectroscopic surveys are incomplete, leaving the intervening cosmic-web relatively unexplored. 

 In order to explore the large-scale structure at the relatively high redshift of these absorption features, we utilized the machine-learning generated WISE-SuperCOSMOS photometric redshift catalog of \cite{bili16}. The catalog, which consists of $>$20 million galaxies, was constructed by jointly analyzing the mid infrared WISE catalog \citep{cutr13} with cross-matched sources in the optical SuperCOSMOS catalog \citep{peac16}. The photometric redshifts were calibrated using an artificial neural-network (ANN) using Galaxy and Mass Assembly (GAMA) spectroscopic data \citep{driv09}, and produce photo-zs with an average accuracy of $\sigma_z = 0.033$. The catalog redshifts rapidly lose completeness beyond $z\approx$~0.3. 

We separated the catalog into redshift bins of $\delta z = 0.04$, which corresponds roughly to the 1$\sigma_z=0.033$ uncertainty of the photometric redshifts. We then created all-sky maps of the galaxy distribution (number of galaxies per pixel) using the Healpix scheme \citep{gors05} with N$_{side}$=256 ($\theta_{pix}\sim$0.229 arcmin),  and the Galactic plane and other contaminated regions masked out \citep{bili16}. In order to highlight any large-scale structure that might be evident in the images, we convolved the map with a FWHM=2.5$^{\circ}$ Gaussian kernel. Figure 1 shows the WISE-SuperCOSMOS maps in the 25$^{\circ}$~x~25$^{\circ}$ field surrounding 1ES 1553+113, plotted over the eight bins that cover the complete redshift range of the survey.  We note that the 2.5$^{\circ}$ Gaussian smoothing corresponds to 11-16~Mpc at the highest redshifts, and highlights only the largest-scale filaments and super-clusters \citep[e.g.,][]{ryu03}. Due to the varying completeness we've normalized the maps in each bin. Using the pixel value of these normalized maps at the location of 1ES~1553+113, the largest intervening structure is at $0.21<z<0.25$, followed by $0.29<z<0.33$. 
\newpage
To explore these structures in more detail, we have re-binned the catalog into $\delta z = 0.02$ slices, and plotted the pixel value along the l.o.s. to 1ES~1553+113 as a function of redshift (Fig. 1 bottom, solid blue line), along with the mean (dashed green line) and rms (shaded blue region) of the pixel values in the 25$^{\circ}$~x~25$^{\circ}$ images after convolution. The highest peak is 5.3$\sigma$ and occurs at z=0.23, followed by the 3.8$\sigma$ peak at z=0.31. 

Also plotted on Fig. 1 are the redshifts of the three WHIM absorption detections of \cite{nica13}. Both the $z_X = 0.312$ and $z_X = 0.237$ absorption lines correspond, within the photometric redshift uncertainty, to a significant peak in the l.o.s. redshift distribution. However, the $<z_X > = 0.133$ does not correspond to a significant over-density. Given the narrowness of the absorption features, $\Delta z \approx 0.001$ is  both cases \citep{nica13}, it is not likely that these are integrated absorptions over the filaments. The filaments instead indicate an increased likelihood for there to be an absorption system along the l.o.s., and warrants further investigation. We have placed the WISE-SuperCOSMOS photometric catalog maps in Healpix format, along with the code to create them, online at \href{https://github.com/sheabrown/WHIM_Filament}{https://github.com/sheabrown/WHIM\_Filament}. 

\begin{figure}[h]
\begin{center}
\includegraphics[scale=0.8,angle=0]{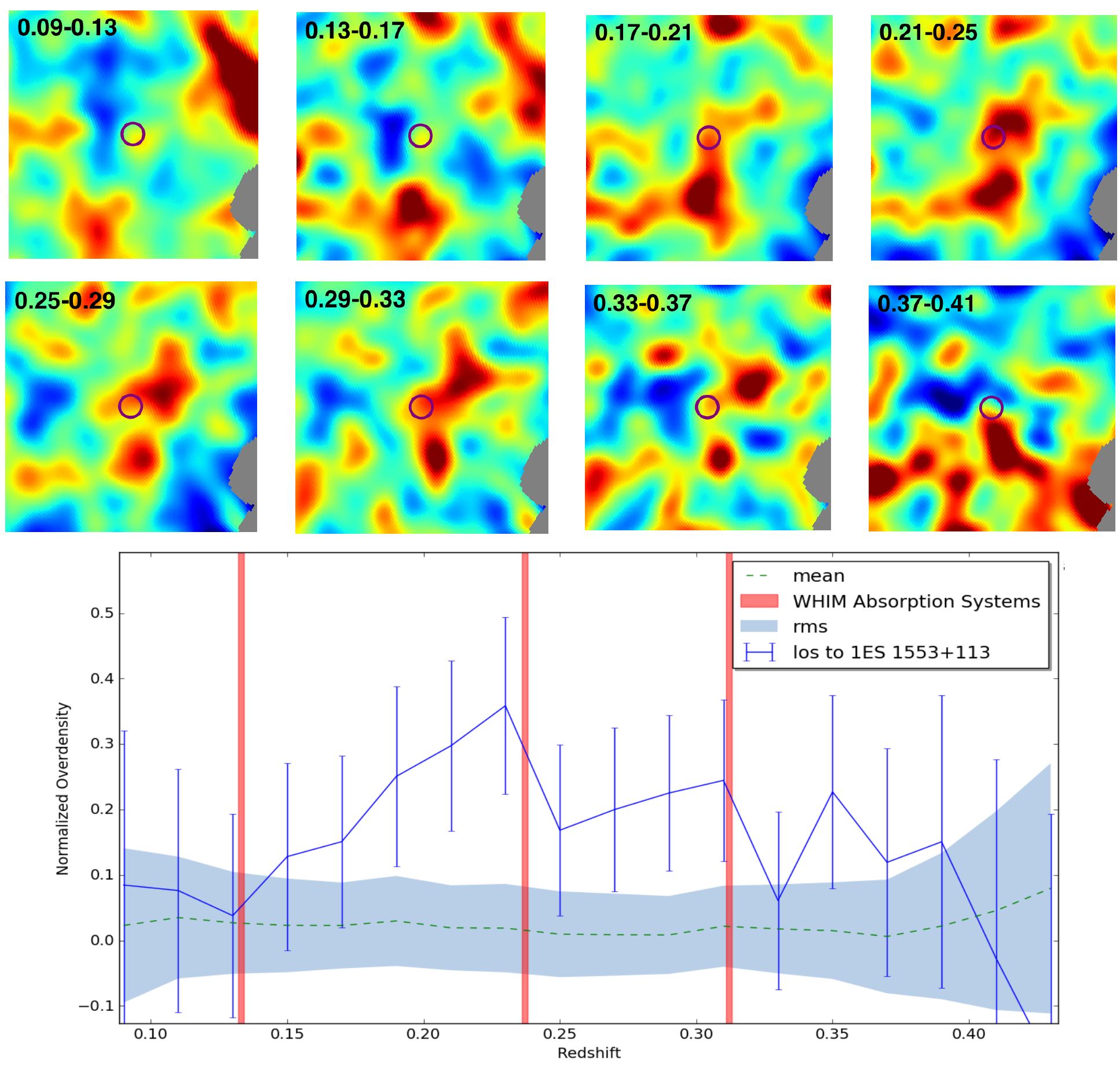}
\caption{\label{fig:1} Top: Sub-images are maps of normalized number-density of galaxies in the 25$^{\circ}$~x~25$^{\circ}$ field surrounding 1ES 1553+113 (dark circles) in the labeled redshift bins. Bottom: Pixel value along the l.o.s. to 1ES~1553+113 as a function of redshift (solid blue line), along with the mean (dashed green line) and rms (shaded blue region) of the pixel values in the 25$^{\circ}$x25$^{\circ}$ images, re-binned to $\delta z = 0.02$.}
\end{center}
\end{figure}



\end{document}